\documentclass{aa}
\usepackage[]{natbib}
\usepackage[]{graphics}
\usepackage[]{lscape}
\bibpunct{(}{)}{;}{a}{}{,}
\newcommand{\etal}{{\it et al.}}

\newcommand{\sbu}{\mbox{mag arcsec$^{-2}$}}
\newcommand{\kms}{\mbox{km~s$^{-1}$}}
\newcommand{\mhi}{\mbox{M$_{HI}$}}
\newcommand{\mmol}{\mbox{M$_{mol}$}}
\newcommand{\md}{\mbox{M$_{dust}$}}

\newcommand{\area}{\mbox{D$^2_{25}$}}
\newcommand{\LB}{\mbox{L$_B$}}
\newcommand{\Lx}{\mbox{L$_X$}}

\begin{document}
\title{A new catalogue of ISM content of normal galaxies.}

\author{D. Bettoni\inst{1}
\and
G. Galletta\inst{2}  
\and
S. Garc\'{\i}a-Burillo\inst{3} }

   \offprints{G. Galletta}

\institute{Osservatorio Astronomico di Padova, Vicolo Osservatorio 5, 35122 Padova, Italy
\and
Dipartimento di Astronomia, Universit\`a di Padova, Vicolo Osservatorio 2, 35122 Padova, Italy
\and
Observatorio Astron\'omico Nacional-OAN, Apartado 1143, 28800 Alcal\'a de Henares- Madrid, Spain}

\date{Received ...; accepted ...}

\authorrunning{Bettoni et al.}
\titlerunning{ISM in normal galaxies}

\abstract{We have compiled a catalogue of the
 gas content for a sample of 1916 galaxies, considered to be a fair
 representation of `normality'. The definition of 'normal' galaxy
 adopted in this work implies that we have purposely excluded from the catalogue
 galaxies having distorted morphology (such as interaction bridges, tails
 or lopsidedness) and/or any signature of peculiar kinematics (such as
 polar rings, counterrotating disks or other decoupled components). In
 contrast, we have included systems hosting active galactic
 nuclei (AGN) in the catalogue. This catalogue revises previous
 compendia on the ISM content of galaxies published by \citet{bregman}
 and \citet{casoli}, and compiles data available in the literature
 from several small samples of galaxies. Masses for warm dust, atomic
 and molecular gas, as well as X-ray luminosities have been converted
 to a uniform distance scale taken from the Catalogue of Principal
 Galaxies (PGC). We have used two different normalization factors to
 explore the variation of the gas content along the Hubble sequence:
 the blue luminosity (\LB) and the square of linear diameter (\area ). 
Our catalogue significantly improves the statistics
 of previous reference catalogues and can be used in future studies to
 define a template ISM content for 'normal' galaxies along the Hubble
 sequence. The catalogue can be accessed on-line and is also available
 at the Centre des Donn\'ees Stellaires (CDS).
\keywords{ catalogues -- galaxies: ISM --  galaxies: general --  
Infrared: galaxies -- Radio lines: galaxies -- X-rays: galaxies 
-- Sub-millimeter}}

\maketitle
 
\section{Introduction}

Different surveys have been conducted thus far to establish if
galaxies can be characterized by a canonical gas content that may vary
significantly along the Hubble sequence (early-type versus late-type
objects) and/or depend on the particular environment (isolated versus
strongly interacting galaxies) \citep{knapp, bregman, casoli,
fabbiano, burstein, beuing}. A global study of the ISM content in
galaxies requires a multi-wavelength approach: far infrared (FIR)
observations at 60 and 100 $\mu$m trace  the warm dust content, while 
those at 140-200 $\mu$m the cold dust; mm lines of carbon monoxide
(CO) and the 21cm line of HI must be used to infer the molecular and
atomic gas content, respectively: the X-ray wave-band is used to
study the hot gas component. However, results obtained from previous
surveys have strong biases due to poor statistics (uneven coverage
along the Hubble sequence) and, also, due to the inclusion of galaxies
which are known to be `peculiar'. As an example, the selection of
candidate galaxies to be observed in CO surveys have been typically
based on Far Infrared Luminosity (FIR) criteria. This often favors the
inclusion of `peculiar' galaxies in surveys and introduces a bias
against `normal' systems.

In this work we present a new catalogue describing the Interstellar
Medium (ISM) content in a sample of 1916 `normal' galaxies. A first
preliminary version of this catalogue was used by Bettoni et
al. (2001) (Paper I) as a reference in their study of the gas content
of gas-accreting galaxies. First of all, this requires the adoption of
a non-obvious definition for `normality'. In this work we will
consider a galaxy to be `peculiar', hence to be excluded from the
catalogue, if it has any sign of perturbed morphology due to a recent
or ongoing interaction like tidal tails, bridges, warps or disk
lopsidedness. Furthermore, with our definition of normal galaxy, we
also exclude from this catalogue any galaxy having any sign of
peculiar kinematics, revealed by the presence of polar rings,
counterrotating cores/disks or other decoupled components. With this
definition, we do not exclude in principle galaxies belonging to
groups or clusters: here, a `normal' galaxy is not necessarily an
`isolated' galaxy, but rather a dynamically relaxed stable
system. Nuclear activity is far from universal among galaxies,
although most galaxies are known to host super-massive black holes in
their nuclei. We purposely include active galaxies in our sample,
considering that activity may represent just an episodic stage in the
evolution of normal galaxies that appear as globally unperturbed, and
therefore fitting our requirements.

The ISM catalogue has been compiled making use of several sources
available in the literature. All values have been re-scaled using a
common distance reference taken from the Catalogue of Principal
Galaxies (PGC). We derive mean values for the ISM content along the
Hubble sequence, using a survival analysis method that properly takes 
into account upper limits. The various gas content estimators have
been normalized using both the galaxy absolute blue luminosity \LB\
and the size \area. The latter allows the direct comparison of our results
with those of previous works that used either \LB\ \citep{bregman} or
\area\  \citep{casoli} to derive canonical mean values for galaxies of
different Hubble types.

The new catalogue significantly improves the statistics of previous
reference catalogues. This is especially relevant in the case of early
type galaxies: compared to previous studies, our sample improves the number 
of early-type galaxies included by a factor $>$3 (from ellipticals to 
lenticulars). This catalogue is intended to be used
in future studies as a reference for the ISM content in normal
galaxies. The catalogue can be accessed on line and is also available
at the Centre des Donn\'ees Stellaires (CDS).

\section{Data compilation}

\subsection{The data search}

We searched in the literature for the different available papers
(both survey papers and those dealing with case-by-case studies 
\citep{knapp, roberts, casoli,fabbiano, burstein, beuing}) and 
collected all the published data on the different tracers
of ISM in galaxies. This produced a preliminary list of more than 
3400 candidate galaxies. From this list, all the systems having an 
acknowledged peculiarity had to be excluded.  In addition, we found 
148 galaxies, included in this preliminary compilation, that are 
listed in the \citet{veron} catalogue of active galaxy nuclei
(AGN).  For the reasons explained above, we decided to include these 
galaxies in our sample, although they will be discussed separately.

\subsection{Discarded types of galaxies}
We considered all the systems belonging to at least one of the following 
categories to be {\it globally peculiar} and we therefore removed them 
from our sample:

1) {\it Interacting and disturbed galaxies}. They are mainly listed in the 
\citet{arp}, \citet{am} and \citet{vv} catalogues and appear to be interacting 
with a close object or have an evident sign of distortion in their
disks/spheroids.  This category includes 1422 galaxies, i.e., the bulk
of the systems we discarded to build this compilation.

2) {\it Polar ring galaxies}. Some galaxies appear to host polar- or
very inclined- rings of gas and/or stars.  In this category of
peculiar galaxies we find ellipticals \citep{bg,hawarden} such as NGC
5128 and disk galaxies \citep{rubin3, whitmore} as NGC 4650A. These
decoupled components betray a passed accretion episode. The origin of
polar rings is connected to an interaction or merging between a galaxy
and a `donor' source which may be another galaxy or intergalactic
gas. The ISM content of polar rings has been discussed in Paper I. We
found 51 polar ring galaxies, which were removed from
our catalogue.

3) {\it Galaxies with counterrotation}. Counterrotating galaxies have
a significant amount of their gas and/or stars
\citep{gall96,bettoni2001} rotating in opposite direction with respect
to the majority of the stars. Similarly to the case of polar rings,
the counterrotation phenomenon is also considered to be linked with
interaction or merging episodes. Sixty galaxies were discarded due to
the presence of stellar- or gas- counterrotating components.

\subsection{The final sample}
After exclusion of all the systems belonging to one of the three
categories enumerated above, the final sample of normal galaxies
includes 1916 objects.  The data presented in this catalogue have been
collected from the following sources:

\begin{itemize}
\item FIR observations, at 60 and 100 $\mu$m from \citet{knapp} catalogue and 
the Lyon-Meudon Extragalactic Database (LEDA) \citep{leda}. These references 
furnish the bulk of the FIR data used here: 992 detections and upper limits from 
\citet{knapp} and 711 from LEDA. In the latter case, both the 60 $\mu$m and 100 $\mu$m 
fluxes (S$_{60}$ and S$_{100}$) are available in the raw data, kindly provided
by G. Paturel. Many of the galaxies (346) are also listed in 
\citet{roberts}. Additional IR data for 81 galaxies have been extracted 
from the NASA/IPAC extragalactic database (NED) and \citet{lavezzi} 
(30 galaxies).

\item CO(1--0) observations for 175 galaxies have been taken from a long 
list of published papers. We extracted data also from \citet{roberts}  (58) and 
from the references listed in \citet{casoli}, including 299 galaxies studied in 
previous works.

\item HI data, 855 from LEDA \citep{leda}, 277 from \citet{roberts}, and 
251 from various sources, including \citep{casoli} and NED.

\item X-ray data from papers based on observations made with the Einstein 
\citep{fabbiano, burstein} and the ROSAT \citep{beuing} missions. They 
furnish  290, 683 and 247 values respectively, with many galaxies in common.
A more recent revision of these data for 343 galaxies has been presented
by \citet{osullivan}.

\end{itemize}

For each galaxy we also collected the distance, the blue luminosity
and the size. To standardize the information contained in our
catalogue we decided to use the morphological classification,
distances, sizes and optical luminosities derived from a single
source: the LEDA archive. These data were used to normalize the values
of FIR and X-ray luminosities as well as the atomic and molecular gas
masses. More precisely, we have extracted the following items for each
galaxy: PGC number, morphological type and related numerical parameter 
$t$, distance modulus and absolute magnitude. The distance moduli are 
mainly derived from redshifts, corrected for Virgocentric inflow and 
adopting H$_o$=70 \kms\ Mpc$^{-1}$.  When redshift was not available, 
we used the photometric distance modulus, if present in LEDA. Information 
lacking in LEDA for some galaxies was completed using the ADS bibliographic 
archive, the NED database or the SIMBAD service of the Strasbourg Centre 
of Donn\'ee Stellaire (CDS). Finally, the very few galaxies having no 
classification in the literature were classified by us looking at the 
optical images extracted from NED.

\begin{figure*}
\resizebox{15cm}{!}{\includegraphics{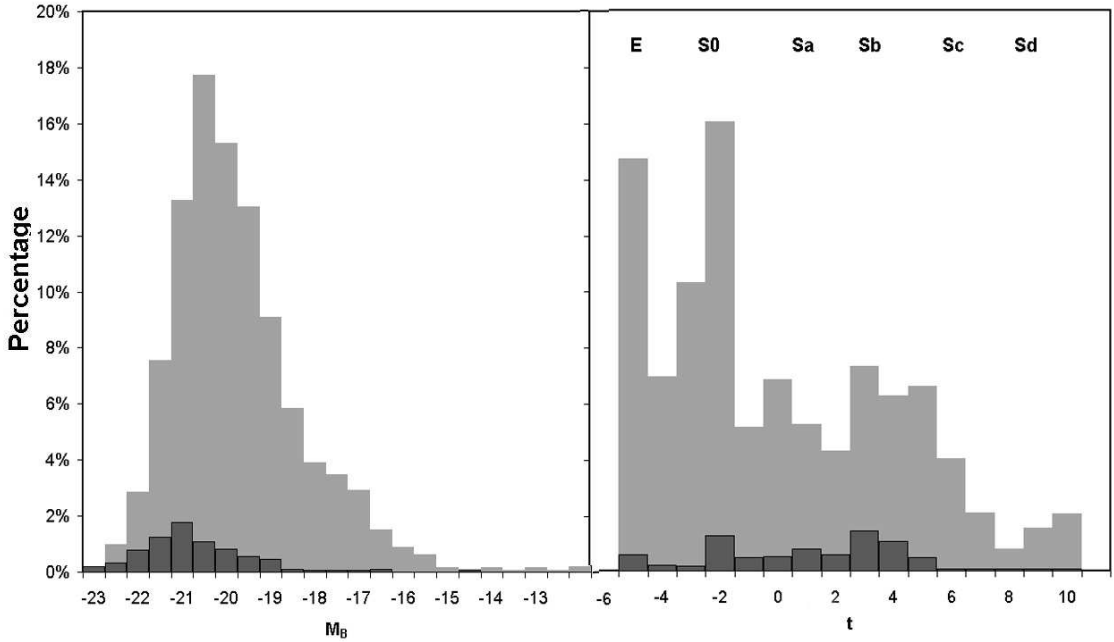}}
\caption{Luminosity function and distribution according to 
morphological type derived for the galaxies in the catalogue. Darker histograms represent AGN hosts,
while light gray bars include all galaxies in the sample.}
\label{lum_funct}
\end{figure*}

The luminosity function and the distribution of morphological types
are shown in Fig.~\ref{lum_funct}. The distribution of morphological
types has two maxima for ellipticals and lenticulars. However, the
rest of morphological types are equally represented in the catalogue
with similar percentages. Altogether, spirals and irregulars (t$\ge$0)
represent 40\% of our sample.  In Fig.~\ref{lum_funct} galaxies
hosting AGN are represented with dark shaded histograms. At first
sight, their mean blue luminosity is slightly above the average value
for the whole sample; on the other hand, the morphological type
distribution of AGN is more weighted by spiral galaxies compared to 
the whole sample.

\section{Data processing}

\subsection{Mass estimates}\label{methods}

We have estimated total gas masses and/or luminosities following
standard procedures (e.g. see \citet{roberts}). The number of
detections and upper limits in our sample are listed in Table
\ref{numbers}. In the following formulae, the distance $d$ is always
in Mpc.

From the value of the IR fluxes at 60 and 100~$\mu$m we estimated the
dust temperature (K):
 
$$ T_d=49 \times (S_{60}/S_{100})^{0.4} $$ 
The corresponding warm dust mass, in solar units, follows from the expression: 
$$Log\ M_d =-2.32+Log\ S_{100}+2\ Log\ d+ Log(exp(\frac{144.06}{T_d})-1) $$ 
Where $S_{100}$ is in mJy. When data for a single galaxy were available in 
several catalogues, we compared the mass values and the data quality, and 
derived accordingly in Table 1 a weighted mean value.

HI gas masses were derived mainly from 21 cm fluxes $S_{21}$ or
from the m$_{21}$ magnitudes of LEDA. From individual papers giving HI masses 
only, we have scaled the published values to the adopted galaxy distance. 
>From m$_{21}$, we used the expression:
 $$Log\ M_{HI} = 5.37-0.4(m_{21c}-17.4)+2\ Log\ d$$ 
while if S$_{21}$, in Jy \kms, was available we used the formula:
$$Log\ M_{HI} = 5.37+\ Log\ S_{21}+2\ Log\ d$$ 
Similarly, we have produced mean values when data from different sources were
simultaneously available. When both upper limits and detections were available
only detections have been used to compute mean values. Moreover, when several upper limits 
were available, only the lowest value has been adopted.

The data needed to derive the molecular gas content have been taken from various sources (see
Appendix A). The formula used to derive molecular gas masses from
CO(1--0) line fluxes ($S_{CO}$ in Jy \kms) is: 
$$Log\ M_{mol} = 4.17 + 2\ Log\ d + Log\ S_{CO}$$
We implicitly assume a constant CO/H$_2$ conversion factor
$\chi$ = $N(H_2)/I_{CO}$= 2.3 $\times$ 10$^{20}$ mol/K km s$^{-1}$
\citep{strong}; $M_{mol}$ includes the helium mass fraction by multiplying the H$_2$ mass by 1.36. 
When only mass values were available in the consulted references, these 
have been scaled to the distances assumed here. 
As discussed in paper I, a source of uncertainty for the estimates of M$_{mol}$ comes from the
undersampling of some of the sources mapped in CO. However the majority of CO observations
considered here sample fairly well the emission in the inner galaxy disks. As we expect the bulk of
CO emission to come precisely from these regions, the derived M$_{mol}$ should be taken as a
meaningful lower limit for the total molecular gas mass. 
A variation of
the CO-H$_2$ conversion factor--$\chi$--can neither be excluded and may be caused by
the different metallicities or physical conditions for molecular gas among the galaxies 
in the sample. Before any further improvement can be envisaged we will assume the above derived
values as a good estimate of the molecular gas masses.

X-ray emission in galaxies may derive from diffuse gas and/or from
discrete sources.  This means that the mass estimate of hot gas strongly
depends on the prevalent mechanism at work in a particular
galaxy. Hence, mass estimates cannot be described by a simple
formula. We thus decided to use the X-ray luminosities, instead of the
masses, to describe the canonical X-ray yield for galaxies in our
sample.  A discussion of the differences existing in literature data 
on X-ray observations with different telescopes has been made by 
\citet{beuing}.

We have 325 detections and 626 upper limits among the inactive galaxies;
and 84 detections and 29 upper limits from galaxies indicated in the
\citet{veron} catalogue as AGN (139) or containing quasars (6) or BL 
Lac (3).

To avoid zero-order biases we have normalized the gas masses and
luminosities using two different normalization factors: the total blue
luminosity--\LB\--and the square of the linear diameter measured at 25
\sbu--\area.  These parameters are available for all the 1916 galaxies
of our sample. The absolute blue magnitudes (from LEDA catalogue) are
corrected for dust absorption and cosmological reddening.  The global
correction is also a function of the type
\citep{leda}; the correction factor increases toward later types.

The mean value of $Log ~\LB/\area$, derived for the galaxies in our sample is: 
$$ Log ~\LB/\area = -0.013\ t + 7.6 $$ with \LB/\area in
L$_\odot$/kpc$^2$ with an rms of $\pm$0.3 (1$\sigma$). As we shall see
in the following paragraphs, this relatively low slope produces similar
trends in the M/\LB ~and M/\area ~mean ratios derived in this work.

\subsection{Stability of the results.}

In principle, M/\LB\ and M/\area\ ratios do not depend at all on the
adopted distance of the galaxy. This makes our results 
insensitive to the specific value of the Hubble constant
adopted. However, distances and magnitudes are affected differently by
the corrections for the Virgocentric motion and by the galactic and
internal absorptions. Hence, M/\LB\ and M/\area\ may change
significantly.  Since the first approach to this catalogue
was published by \citet{bettoni2001}, the LEDA database has been
modified, because the stored data are continuously being updated.  In
addition to these changes, a significant amount of new data from other
sources have been included in the new catalogue.

While these changes forced us to recalculate all the ratios, this
allowed us to test the stability of our results. Comparing the data
extracted from this database in different dates in the last two years,
we found that the 90\% of the Log M/\LB\ and the 98\% of the Log
M/\area\ values calculated from these different data extractions
differ by less than 1$\sigma$, with no variation larger than
2$\sigma$. Within these ranges we expect our conclusions to be stable.

\subsection{Derivation of mean ratios: Log~$M/D^2$--Log~$M/L$}

In order to define the mean content of each tracer according to the
type, we made use of survival analysis methods
\citep[see][]{feigelson}, applied to the different ensembles of
M/L$_B$ and M/\area\ data. This analysis tool properly takes into
account both detections and upper limits (UL) in order to derive
representative averages. The mean values are derived and shown in the
following Tables, binned according to the morphological type
code. This method also allows a more direct comparison of our data
with that from literature.

Survival analysis \citep{feigelson} programs are part of several
packages of data analysis, such as IRAF, and may be applied when a
number of upper or lower limit values are available together with
values from detections. Making use of the Kaplan-Meier
estimator, we derived, for each $t$,  mean values of mass/light
and mass/area ratios and the corresponding standard deviations.  
When all the galaxies considered within an interval of $t$ are
detected, these values coincide with that from arithmetic
means. On the contrary, the method cannot be applied when both upper
and lower limits are present, for instance when considering the mass
ratios (e.g. \mmol/\mhi). In such a case, we present ratios of mean
values.

\subsection{Description of the full version of the catalogue.}

All the data of this catalogue are listed in Table 1, which is only
available in digital form. The first page is shown in Table 1 of this
paper, as an illustration of the on-line catalogue format.  The
columns are: (1) the PGC identifier and (2) the galaxy name, (3) the
numerical type code $t$ \citep{rc3,leda}, (4) this column indicates
the presence of nuclear activity \citep{veron} with the letter `a',
(5) the adopted distance $d$ (in Mpc), (6) the logarithm of the total
blue luminosity \LB\ in solar units, (7) the logarithm of the linear
size $D_{25}$ in kpc corresponding to the 25 mag arcsec$^{-2}$
isophote.  The values for each tracer (dust, HI, molecular gas, X-ray)
span four columns: the first two are mass vs. \LB\ and the next two
are mass vs. area, defined as \area. Values corresponding to
detections of a galaxy in the corresponding tracers are in the first
and in the third columns while upper limits are in the second and
fourth columns.  All masses and luminosities are in solar units.
Following this scheme, the following columns contain: Log \md/\LB\
detections (col. 8) and upper limits (col. 9), Log \md/\area\
detections (col.  10) and upper limits (col. 11), Log \mhi/\LB\
detections (col. 12) and upper limits (col. 13), Log \mhi/\area\
detections (col. 14) and upper limits (col.  15), Log \mmol/\LB\
detections (col. 16) and upper limits (col. 17), Log
\mmol/\area\  detections (col. 18) and upper limits (col. 19), Log \Lx/\LB\ 
detections (col. 20) and upper limits (col. 21), Log \Lx/\area\  detections
(col. 22) and upper limits (col. 23).
References for the gas masses are listed in the last column (24). The 
reference codes are given in the Appendix.

This table contains all the essential information compiled for this
catalogue.  This compilation is intended to serve as a template for
future studies on the gas content in selected types of galaxies.

\section{Analysis of main results}

Tables \ref{numbers}--\ref{LX_t} give the normalized mean gas
contents--Log~$M/D^2$ and Log~$M/L_B$-- and the detection rates for the
galaxies of the sample, as a function of the Hubble type.  The
galaxies are identified according to the numerical code $t$ (RC3
morphological type). Some types have two numerical codes: elliptical
galaxies span the types -5 and -4; S0-a the types -1 and 0 and Sc
galaxies fill the types 5 and 6; all the remaining galaxy types are
identified by a single code. Note that the code 5.00 also includes
spiral galaxies with ambiguous classification (type S?). In our sample
there are only 7 galaxies falling into this category.

\subsection{Detection rates}
As far as detection rates are concerned, we have found the following
trends:
\begin{itemize}

\item
The detection rate of elliptical galaxies in X-ray wavelengths from
our data is 1.3--4 times higher than that of late-type galaxies. As
expected, the percentage of X-ray detections for spirals increases by
10\% if AGN galaxies are included in the sample.

\item 
On the other hand, compared with ellipticals, both CO and HI
detections are $\sim$4 times more common in disk galaxies (note
however that $\sim$40\% of type t=-5 ellipticals have been
detected in CO or HI).  This high detection rate in spirals reflects the
higher cold gas content of disk galaxies.  The dust detection roughly
follows the HI detection rates.

\end{itemize}

\subsection{Comparison with previous work}

\citet{bregman} and \citet{casoli} have previously addressed the study 
of the ISM content in galaxies and its variation along the Hubble
sequence. Both groups based their study on two samples of 467
\citep{bregman} and 582 \citep{casoli} galaxies, compiled and processed 
according to different criteria. Our catalogue improves by a significant 
factor ($\sim$3--4) the overall statistics of previous samples. The most
significant improvement corresponds to the left branch of the Hubble
sequence: compared to previous works, the number of early-type
galaxies (t $<0$) with firm detections processed for this
catalogue is $>$3 times larger.

\begin{figure*}

\resizebox{15cm}{!}{\includegraphics{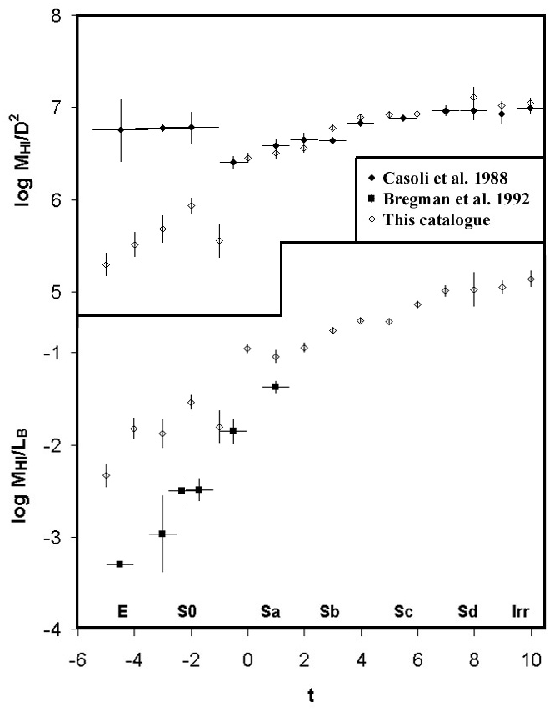}\includegraphics{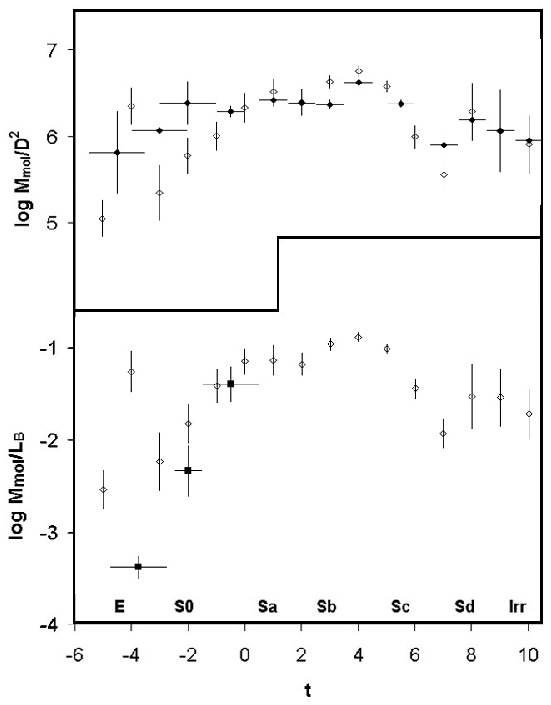}}
\caption{Comparison between the mean Log M/L ratios for HI (left panels) and for molecular gas 
(right panels) extracted from the present catalogue and mean Log M/L ratios of previous studies 
\citep{bregman,casoli}. In the upper panel the mean masses are normalized with respect to \area, 
as adopted by \citet{casoli}, while in the lower panel \LB\ is the normalization factor, as 
adopted by \citet{bregman}.  Our estimates agree with those published by \citet{casoli} only for
types later than S0.}

\label{comparison}
\end{figure*}

\citet{bregman} analyzed their sample using other non-parametric 
tests, based on rank, 
to study the variation of the mean gas content with morphological
type. The estimated averages were normalized to the galaxy blue
luminosity.  Their results indicate that the mean content of atomic
gas (in M/L$_B$ scale) decreases by two orders of magnitude from
early-type spirals (Sa) to ellipticals (E). A similar result is
obtained for the mean content of molecular gas. Altogether, these
results underlined that ellipticals have a neutral gas content
significantly lower than lenticulars (see Fig.~\ref{comparison}). The
sample of \citet{bregman} has only 7 (35) firm detections in CO (HI)
for galaxies spanning the lenticular-elliptical range.

The scenario depicted by \citet{casoli} seems at odds with
\citet{bregman}'s results, however.
\citet{casoli} used a survival analysis method and 
normalized their data by \area. 
 According to \citet{casoli}, the atomic mean gas content shows little
 variation going from elliptical to lenticular galaxies (see upper
 panel of Fig.~\ref{comparison}).  Within the same type range, the
 molecular gas content varies by less than one order of magnitude: i.e.,
 significantly less than shown by \citet{bregman}. \citet{casoli}
 excluded from their sample galaxies in cluster environments, tagged
 as HI-deficient. This might explain why, in contrast to
 \citet{bregman}, \citet{casoli} find no significant trend for the HI
 content.  The sample of
\citet{casoli} has 11 (13) firm detections in CO (HI) for galaxies spanning the
lenticular-elliptical range.

These two scenarios might only be reconciled if the mean L$_B$/\area\
ratios were seen to strongly depend on morphological type in the
`conflicting' range: ellipticals would need to be much fainter, by at
least two orders of magnitude, with respect to S0 of comparable
sizes. As it has been discussed in Sect.~\ref{methods}, the estimated
trend of $Log \LB/\area$ is much smoother, however; this is confirmed
by the similar trends obtained using either M/\LB  or M/\area  as gas
content indicators.

The strong disagreement between these two studies should be
attributable to their poor statistics.  This is probably the reason
why our estimates of the neutral gas content match neither of previous
claims in this range. As it is shown in Table \ref{numbers}, the
present catalogue is based on 31 CO detections and 132 HI detections
within the range of early type galaxies (ellipticals to
lenticulars). Our results, normalized with respect to \LB\ and \area\
both indicate a similar trend: a moderate one order of magnitude
decrease of the mean neutral gas content (atomic and/or molecular)
going from S0 to E galaxies.

\begin{figure}
\resizebox{9cm}{!}{\includegraphics{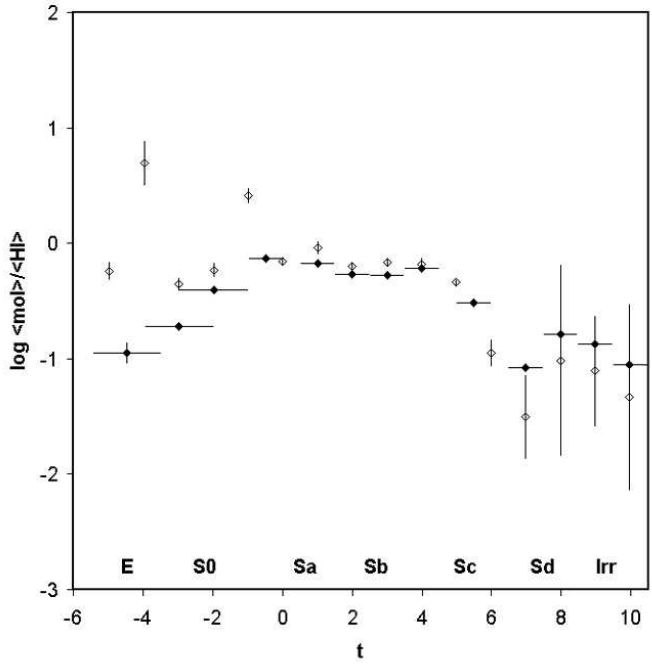}}
\caption{The mean molecular to atomic gas content ratio as a function of Hubble type. The open
symbols are derived from our catalogue, while full symbols
represent ratios published by \citet[][full symbols]{casoli}.}
\label{mol_HI}
\end{figure}

It is worth noting that, in contrast to \citet{casoli}, we found that
the \mmol/\mhi\ ratio is roughly constant along the Hubble sequence,
except for the latest type systems (Sd and Irr) where this ratio
decreases by one order of magnitude. Early-type galaxies and early
spirals possess $\sim$40\% of their neutral gas content in the
molecular phase, while late-type spirals and irregulars have
$\sim$90\% of their neutral gas content in atomic form. In
Fig.~\ref{mol_HI}, we plotted our data together with values derived by
\citet{casoli}.

\subsection{AGN versus non-active galaxies}

We have found no significant differences in the neutral gas content
between galaxies hosting AGNs and those considered as non-active. For
this reason, the mean data in Tables \ref{Tdust}-\ref{Thi} do not
distinguish between active and inactive galaxies, even if it
is possible to identify both categories of galaxies in our catalogue.

\begin{figure}
\resizebox{9cm}{!}{\includegraphics{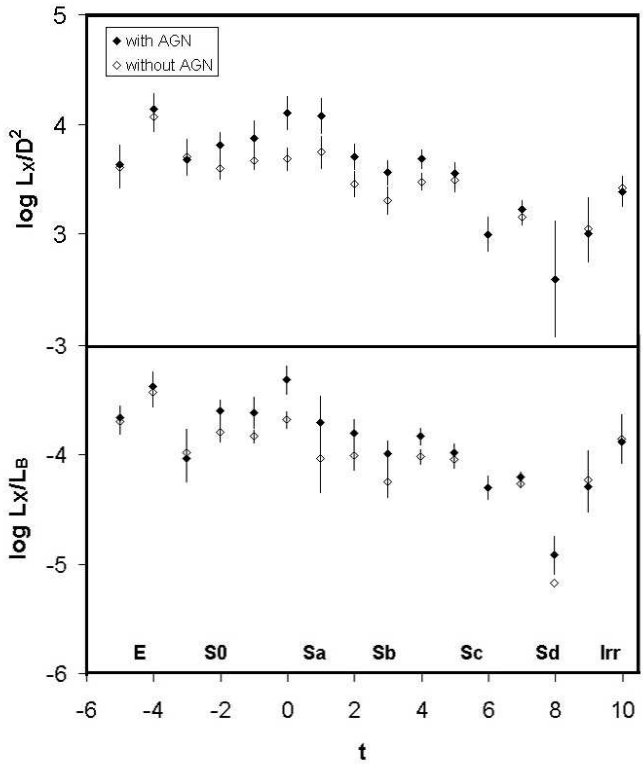}}
\caption{Variation of mean Log L$_X$/\area\ and Log L$_X$/L$_B$ ratios either including
or excluding AGN host galaxies. }
\label{AGN_NOAGN}
\end{figure}

In our database there are 20 active galaxies that are more luminous than 
Log \Lx/\LB=-2, while most of the remaining galaxies have values between 
-2 and -4. Apart from the overluminous galaxies, the mean ratios \Lx/\LB\ and
\Lx/\area\ show an almost flat trend going from elliptical galaxies 
to Sa and a slowly decreasing trend for later types (Fig.~\ref{AGN_NOAGN}).  
In Table \ref{LX_t} we publish the mean values obtained by both including and 
excluding AGN.

\section{On-line access}

We have presented a catalogue devoted to study the ISM content in a
sample of normal galaxies. The catalogue has been compiled based on
previous catalogues with the addition of a significant volume of new
data. This is intended to serve as a reference defining the canonical
gas content of 'normal' galaxies along the Hubble sequence. The latest
version of the catalogue is available at
http://dipastro.pd.astro.it/galletta/ismcat/.  The number of galaxies
included, 1916 systems, is 3--4 times higher than previous studies,
and fills a gap currently existing in the ISM studies of cold gas for
early-type galaxies.  All the catalogue data have been homogenized and
reduced to the same distance scale and can be accessed as a single
text list, formatted in columns, or consulted at CDS. Upon request, it
can be obtained from the authors as an Excel-formatted database.

\section{Acknowledgments}
The authors thanks Dr. D. Hogg, for the useful comments to this work.

This research made use of Vizier service \citep{vizier}, the 
SIMBAD database (operated at CDS, Strasbourg, France), the NASA/IPAC 
Extragalactic Database (NED) (which is operated by JPL, California 
Institute of Technology, under contract to NASA) and 
NASA's Astrophysics Data System Abstract Service (mirrored in CDS of
Strasbourg). The authors would like to thank Dr. G. Paturel for kindly 
making the LEDA database's FIR raw data available and Dr. F. Ochsenbein 
for the changes made to the Vizier's query form on our request. 
GG made use of funds from the University of Padova (Fondi 60\%-2002).

\section{Appendix A: References to Table 1}
\begin{tabular}{lll}
\hline
reference & Source of data & ISM \\
code  &                    & tracer \\
\hline
 1a  &  \citet{andreani}  &  HI   \\
 1b  &  \citet{andreani}  &  CO  \\
 2  &  \citet{beuing}     &  X  \\
 3  &  \citet{boselli}    &  CO  \\
 4  &  \citet{bregman}    &  CO  \\
 5  &  \citet{burstein}   &  X  \\
 6a  &  \citet{casoli91}  &  HI  \\
 6b  &  \citet{casoli91}  &  CO  \\
 7  &  \citet{casoli}     &  CO  \\
 8  &  \citet{fabbiano}   &  X  \\
 9  &  \citet{gerin}      &  CO  \\
10a  &  \citet{horellou}  &  HI  \\
10b  &  \citet{horellou}  &  CO  \\
11  &  \citet{knapp}      &  IR  \\
12a  &  \citet{lavezzi}      &  IR  \\
12b  &  \citet{lavezzi}      &  HI  \\
12c  &  \citet{lavezzi}      &  CO  \\
13a  &  \citet{leda}      &  IR  \\
14a  &  NASA/IPAC Extragalactic Database (NED)      &  IR  \\
14b  &  NASA/IPAC Extragalactic Database (NED)      &  HI  \\
14c  &  NASA/IPAC Extragalactic Database (NED)      &  CO  \\
15   &  \citet{nishiyama}      &  CO  \\
16   &  \citet{osullivan}      &  CO  \\
17a  &  \citet{roberts}   &  IR  \\
17b  &  \citet{roberts}   &  HI  \\
17c  &  \citet{roberts}   &  CO  \\
17d  &  \citet{roberts}   &  X  \\
18a  &  Sage 1993         &  HI  \\
18b  &  Sage 1993         &  CO  \\
19  &  \citet{vandriel}   &  HI  \\
20a  &  \citet{welch}    &  HI  \\
20b  &  \citet{welch}    &  CO  \\
21  &  \citet{wiklind}    &  CO  \\
22  &  \citet{fcrao}      &  CO  \\
\hline
\end{tabular}


\begin{landscape}
\begin{table}                                              
{\tiny
\tabcolsep 5pt
\caption{An illustration of the catalogue format.}                                              
\begin{tabular}{rcrrrrrrrrrrrrrrrrrrrrrr} 
\hline
(1)	&		(2)	&	(3)	&	(4)	&	(5)	&	(6)	&	(7)	&	(8)	&	(9)	&	(10)	&	(11)	&	(12)	&	(13)	&	(14)	&	(15)	&
(16)	&	(17)	&	(18)	&	(19)	&	(20)	&	(21)	&	(22)	&	(23)	&	(24)	\\
    101	&	    NGC7803    	&	0.04	&	*	&	  76.98	&	10.49	&	1.23	&	-3.77	&	  *  	&	4.27	&	  *  	&	-0.74
&	  *  	&	7.30	&	  *  	&	  *  	&	  *  	&	  *  	&	  *  	&	  *  	&	  *  	&	  *  	&	  *  	&	11,12a,13b
\\ 
    218	&	    NGC7814    	&	2.02	&	*	&	  15.89	&	10.16	&	1.41	&	-4.18	&	  *  	&	3.16	&	  *  	&	-1.04
&	  *  	&	6.30	&	  *  	&	  *  	&	-1.11	&	  *  	&	6.23	&	  *  	&	  *  	&	  *  	&	  *  	&	13b,14a,22
\\ 
    279	&	    NGC7817    	&	4.10	&	*	&	  34.14	&	10.51	&	1.53	&	-3.59	&	  *  	&	3.86	&	  *  	&	-0.83
&	  *  	&	6.62	&	  *  	&	-0.76	&	  *  	&	6.68	&	  *  	&	  *  	&	  *  	&	  *  	&	  *  	&	12a,13b,22
\\ 
    307	&	    NGC7820    	&	-0.07	&	*	&	  43.77	&	10.14	&	1.28	&	  *  	&	-5.17	&	  *  	&	2.40	&	  *  
&	  *  	&	  *  	&	  *  	&	  *  	&	  *  	&	  *  	&	  *  	&	  *  	&	  *  	&	  *  	&	  *  	&	11 \\ 
    322	&	   ESO409-12   	&	-4.29	&	*	&	 112.05	&	10.65	&	1.62	&	  *  	&	-4.78	&	  *  	&	2.62	&	  *  
&	  *  	&	  *  	&	  *  	&	  *  	&	  *  	&	  *  	&	  *  	&	  *  	&	  *  	&	  *  	&	  *  	&	11 \\ 
    354	&	    NGC7824    	&	2.00	&	*	&	  87.78	&	10.54	&	1.61	&	  *  	&	  *  	&	  *  	&	  *  	&	  *  
&	  *  	&	  *  	&	  *  	&	  *  	&	  *  	&	  *  	&	  *  	&	  *  	&	-2.79	&	  *  	&	4.54	&	16 \\ 
    621	&	   ESO349-31   	&	9.98	&	*	&	   3.16	&	7.22	&	0.00	&	  *  	&	  *  	&	  *  	&	  *  	&	-0.10
&	  *  	&	7.12	&	  *  	&	  *  	&	  *  	&	  *  	&	  *  	&	  *  	&	-2.41	&	  *  	&	4.81	&	5,13b \\ 
    658	&	     UGC81     	&	3.03	&	*	&	  95.81	&	10.44	&	1.53	&	  *  	&	  *  	&	  *  	&	  *  	&	  *  
&	  *  	&	  *  	&	  *  	&	  *  	&	  *  	&	  *  	&	  *  	&	  *  	&	-2.74	&	  *  	&	4.65	&	5 \\ 
    660	&	     NGC16     	&	-2.69	&	*	&	  45.06	&	10.35	&	1.39	&	  *  	&	-5.62	&	  *  	&	1.95	&	  *  
&	-1.87	&	  *  	&	5.70	&	  *  	&	  *  	&	  *  	&	  *  	&	  *  	&	  *  	&	  *  	&	  *  	&	11,17a,17b
\\ 
    684	&	    IC1531     	&	-2.61	&	*	&	 107.35	&	10.80	&	1.69	&	  *  	&	  *  	&	  *  	&	  *  	&	  *  
&	  *  	&	  *  	&	  *  	&	  *  	&	  *  	&	  *  	&	  *  	&	-2.80	&	  *  	&	4.62	&	  *  	&	2,16 \\ 
    698	&	     NGC23     	&	1.20	&	*	&	  66.68	&	10.78	&	1.56	&	-3.64	&	  *  	&	4.03	&	  *  	&	-0.83
&	  *  	&	6.84	&	  *  	&	-0.51	&	  *  	&	7.16	&	  *  	&	  *  	&	  *  	&	  *  	&	  *  	&	12a,13b,22
\\ 
    775	&	   ESO349-37   	&	0.11	&	*	&	 210.38	&	10.85	&	1.74	&	  *  	&	  *  	&	  *  	&	  *  	&	  *  
&	  *  	&	  *  	&	  *  	&	  *  	&	  *  	&	  *  	&	  *  	&	  *  	&	-2.30	&	  *  	&	5.07	&	5 \\ 
    875	&	     NGC43     	&	-1.97	&	*	&	  70.15	&	10.46	&	1.48	&	  *  	&	-4.74	&	  *  	&	2.77	&	  *  
&	  *  	&	  *  	&	  *  	&	  *  	&	  *  	&	  *  	&	  *  	&	  *  	&	  *  	&	  *  	&	  *  	&	11 \\ 
   1030	&	    UGC144     	&	4.00	&	*	&	  81.10	&	10.16	&	1.38	&	  *  	&	  *  	&	  *  	&	  *  	&
-1.35	&	  *  	&	6.05	&	  *  	&	  *  	&	  *  	&	  *  	&	  *  	&	  *  	&	-2.67	&	  *  	&	4.74	&	5,13b
\\ 
   1037	&	     NGC57     	&	-4.90	&	*	&	  78.38	&	11.00	&	1.71	&	  *  	&	-5.29	&	  *  	&	2.30	&	  *  
&	  *  	&	  *  	&	  *  	&	  *  	&	  *  	&	  *  	&	  *  	&	-2.70	&	  *  	&	4.88	&	  *  	&	2,11,16 \\ 
   1160	&	     NGC63     	&	4.90	&	*	&	  17.10	&	9.75	&	0.93	&	-4.46	&	  *  	&	3.43	&	  *  	&	-1.17
&	  *  	&	6.72	&	  *  	&	  *  	&	  *  	&	  *  	&	  *  	&	  *  	&	  *  	&	  *  	&	  *  	&	11,12a,13b
\\
   1351	&	     NGC80     	&	-2.84	&	*	&	  83.21	&	10.84	&	1.58	&	  *  	&	-4.61	&	  *  	&	3.07	&	  *  
&	-1.72	&	  *  	&	5.95	&	  *  	&	  *  	&	  *  	&	  *  	&	  *  	&	  *  	&	  *  	&	  *  	&	11,14b \\
   1362	&	    IC1544     	&	5.30	&	*	&	  82.87	&	10.33	&	1.45	&	  *  	&	-2.73	&	  *  	&	4.69	&	-0.44
&	  *  	&	6.99	&	  *  	&	  *  	&	  *  	&	  *  	&	  *  	&	  *  	&	-2.76	&	  *  	&	4.66	&	5,13b,14a
\\ 
   1371	&	     NGC83     	&	-4.82	&	*	&	  92.77	&	10.79	&	1.53	&	-3.38	&	  *  	&	4.34	&	  *  	&	  *  
&	  *  	&	  *  	&	  *  	&	-1.39	&	  *  	&	6.33	&	  *  	&	  *  	&	  *  	&	  *  	&	  *  	&	11,21 \\ 
   1442	&	     NGC97     	&	-4.85	&	*	&	  72.24	&	10.55	&	1.45	&	  *  	&	-5.01	&	  *  	&	2.64	&	  *  
&	  *  	&	  *  	&	  *  	&	  *  	&	  *  	&	  *  	&	  *  	&	  *  	&	  *  	&	  *  	&	  *  	&	11 \\ 
   1619	&	    NGC108     	&	-1.25	&	*	&	  69.47	&	10.49	&	1.58	&	-4.20	&	  *  	&	3.13	&	  *  	&
-1.00	&	  *  	&	6.33	&	  *  	&	  *  	&	  *  	&	  *  	&	  *  	&	  *  	&	  *  	&	  *  	&	  *  	&
11,14b \\ 
   1772	&	    NGC125     	&	-1.14	&	*	&	  75.58	&	10.67	&	1.55	&	  *  	&	  *  	&	  *  	&	  *  	&
-1.54	&	  *  	&	6.02	&	  *  	&	  *  	&	  *  	&	  *  	&	  *  	&	  *  	&	-2.79	&	  *  	&	4.77	&	8,13b
\\ 
   1787	&	    NGC127     	&	-2.02	&	*	&	  57.68	&	9.49	&	1.19	&	-3.38	&	  *  	&	3.72	&	  *  	&	  *  
&	  *  	&	  *  	&	  *  	&	  *  	&	  *  	&	  *  	&	  *  	&	  *  	&	-1.85	&	  *  	&	5.26	&	8,12a,16
\\ 
   1794	&	    NGC130     	&	-3.04	&	*	&	  64.00	&	9.71	&	1.12	&	  *  	&	  *  	&	  *  	&	  *  	&	  *  
&	  *  	&	  *  	&	  *  	&	  *  	&	  *  	&	  *  	&	  *  	&	  *  	&	-1.97	&	  *  	&	5.50	&	8,16 \\ 
   1816	&	   ESO194-21   	&	-2.96	&	*	&	  44.08	&	9.76	&	1.22	&	-5.19	&	  *  	&	2.12	&	  *  	&	  *  
&	  *  	&	  *  	&	  *  	&	  *  	&	  *  	&	  *  	&	  *  	&	  *  	&	  *  	&	  *  	&	  *  	&	11 \\ 
   1869	&	    UGC305     	&	5.27	&	*	&	 142.43	&	10.67	&	1.73	&	  *  	&	  *  	&	  *  	&	  *  	&	  *  
&	  *  	&	  *  	&	  *  	&	  *  	&	  *  	&	  *  	&	  *  	&	  *  	&	-2.64	&	  *  	&	4.56	&	5 \\ 
   1901	&	    NGC142     	&	3.01	&	*	&	 113.45	&	10.64	&	1.53	&	-3.60	&	  *  	&	3.96	&	  *  	&	-0.91
&	  *  	&	6.66	&	  *  	&	-0.48	&	  *  	&	7.09	&	  *  	&	  *  	&	  *  	&	  *  	&	  *  	&
1a,1b,12a,13b \\ 
   1977	&	    IC1553     	&	5.45	&	*	&	  39.68	&	9.94	&	1.18	&	-3.70	&	  *  	&	3.88	&	  *  	&	-0.78
&	  *  	&	6.81	&	  *  	&	-1.46	&	  *  	&	6.13	&	  *  	&	  *  	&	  *  	&	  *  	&	  *  	&
1a,1b,12a,13b \\ 
   2004	&	    NGC147     	&	-4.60	&	*	&	    .62	&	7.88	&	0.37	&	-5.74	&	  *  	&	1.39	&	  *  	&	  *  
&	-1.72	&	  *  	&	5.41	&	  *  	&	  *  	&	  *  	&	  *  	&	  *  	&	-4.32	&	  *  	&	2.81	&
2,11,12a,16,17a,17b \\ 
   2053	&	    NGC148     	&	-2.10	&	*	&	  19.16	&	9.50	&	1.04	&	  *  	&	-5.25	&	  *  	&	2.17	&	-0.37
&	  *  	&	7.05	&	  *  	&	  *  	&	  *  	&	  *  	&	  *  	&	  *  	&	  *  	&	  *  	&	  *  	&
11,13b,17a,17b \\ 
   2081	&	    NGC157     	&	4.02	&	a	&	  22.70	&	10.70	&	1.44	&	-3.78	&	  *  	&	4.05	&	  *  	&	-0.79
&	  *  	&	7.04	&	  *  	&	-1.12	&	  *  	&	6.71	&	  *  	&	  *  	&	  *  	&	  *  	&	  *  	&	12a,13b,22
\\ 
   2154	&	    NGC160     	&	-0.39	&	*	&	  76.28	&	10.61	&	1.71	&	-4.18	&	  *  	&	3.02	&	  *  	&
-0.83	&	  *  	&	6.37	&	  *  	&	  *  	&	  *  	&	  *  	&	  *  	&	  *  	&	  *  	&	  *  	&	  *  	&
11,13b \\ 
   2228	&	    NGC172     	&	4.01	&	*	&	  41.55	&	10.16	&	1.39	&	-4.53	&	  *  	&	2.85	&	  *  	&	-0.19
&	  *  	&	7.20	&	  *  	&	  *  	&	  *  	&	  *  	&	  *  	&	  *  	&	-3.17	&	  *  	&	4.22	&	5,12a,13b
\\ 
   2253	&	    NGC179     	&	-3.14	&	*	&	  84.14	&	10.40	&	1.39	&	-4.65	&	  *  	&	2.97	&	  *  	&	  *  
&	  *  	&	  *  	&	  *  	&	  *  	&	  *  	&	  *  	&	  *  	&	  *  	&	  *  	&	  *  	&	  *  	&	11 \\ 
   2298	&	    NGC183     	&	-4.90	&	*	&	  77.41	&	10.52	&	1.48	&	-4.60	&	  *  	&	2.96	&	  *  	&	  *  
&	  *  	&	  *  	&	  *  	&	  *  	&	  *  	&	  *  	&	  *  	&	  *  	&	-3.01	&	  *  	&	4.55	&	5,11 \\ 
   2329	&	    NGC185     	&	-4.60	&	a	&	    .59	&	8.01	&	0.39	&	-5.14	&	  *  	&	2.10	&	  *  	&	-2.96
&	  *  	&	4.28	&	  *  	&	-2.87	&	  *  	&	4.37	&	  *  	&	  *  	&	-4.54	&	  *  	&	2.69	&
2,11,12a,13b,16,17a,17b,17c \\ 
   2362	&	    NGC194     	&	-4.86	&	*	&	  73.55	&	10.69	&	1.44	&	-5.09	&	  *  	&	2.73	&	  *  	&	  *  
&	  *  	&	  *  	&	  *  	&	  *  	&	-1.44	&	  *  	&	6.38	&	  *  	&	  *  	&	  *  	&	  *  	&	3,11 \\ 
   2381	&	    UGC412     	&	-0.07	&	*	&	  79.98	&	9.99	&	1.36	&	  *  	&	  *  	&	  *  	&	  *  	&
-0.75	&	  *  	&	6.53	&	  *  	&	  *  	&	  *  	&	  *  	&	  *  	&	  *  	&	-2.43	&	  *  	&	4.84	&	5,13b
\\ 
   2429	&	    NGC205     	&	-4.66	&	*	&	    .32	&	7.80	&	0.24	&	-5.02	&	  *  	&	2.30	&	  *  	&	-2.76
&	  *  	&	4.56	&	  *  	&	-2.83	&	  *  	&	4.49	&	  *  	&	  *  	&	-4.76	&	  *  	&	2.57	&
2,8,11,12a,13b,16,17a,17b,17c,17d \\ 
   2478	&	    NGC216     	&	-1.63	&	*	&	  20.34	&	9.30	&	1.03	&	-3.81	&	  *  	&	3.43	&	  *  	&	0.30
&	  *  	&	7.54	&	  *  	&	  *  	&	  *  	&	  *  	&	  *  	&	  *  	&	  *  	&	  *  	&	  *  	&	11,13b \\ 
   2504	&	    UGC442     	&	6.50	&	*	&	  72.21	&	9.86	&	1.41	&	  *  	&	  *  	&	  *  	&	  *  	&	-0.09
&	  *  	&	6.96	&	  *  	&	  *  	&	  *  	&	  *  	&	  *  	&	  *  	&	-2.59	&	  *  	&	4.45	&	5,13b \\ 
   2540	&	    UGC446     	&	9.93	&	*	&	  91.29	&	9.74	&	1.46	&	  *  	&	  *  	&	  *  	&	  *  	&	0.25
&	  *  	&	7.06	&	  *  	&	  *  	&	  *  	&	  *  	&	  *  	&	  *  	&	-2.28	&	  *  	&	4.53	&	5,13b \\ 
   2547	&	    NGC227     	&	-3.61	&	*	&	  75.61	&	10.74	&	1.59	&	  *  	&	-5.20	&	  *  	&	2.35	&	  *  
&	-0.70	&	  *  	&	6.85	&	  *  	&	  *  	&	  *  	&	  *  	&	  *  	&	-3.28	&	  *  	&	4.28	&
5,11,16,17a,17b,17d \\ 
   2557	&	    NGC224     	&	3.04	&	a	&	    .98	&	10.88	&	1.72	&	-4.23	&	  *  	&	3.21	&	  *  	&	-1.00
&	  *  	&	6.44	&	  *  	&	  *  	&	  *  	&	  *  	&	  *  	&	-4.62	&	  *  	&	2.82	&	  *  	&	8,12a,13b
\\ 
   2604	&	    NGC233     	&	-5.00	&	*	&	  79.40	&	10.34	&	1.53	&	  *  	&	-5.00	&	  *  	&	2.28	&	  *  
&	  *  	&	  *  	&	  *  	&	  *  	&	  *  	&	  *  	&	  *  	&	  *  	&	  *  	&	  *  	&	  *  	&	11 \\ 
   2683	&	   ESO540-19   	&	3.40	&	*	&	  57.07	&	9.96	&	1.21	&	  *  	&	  *  	&	  *  	&	  *  	&	  *  
&	  *  	&	  *  	&	  *  	&	  *  	&	  *  	&	  *  	&	  *  	&	  *  	&	-2.92	&	  *  	&	4.62	&	5 \\ 
   2758	&	    NGC247     	&	6.95	&	*	&	   2.83	&	9.74	&	1.23	&	-4.63	&	  *  	&	2.64	&	  *  	&	-0.61
&	  *  	&	6.67	&	  *  	&	  *  	&	  *  	&	  *  	&	  *  	&	-4.21	&	  *  	&	3.07	&	  *  	&	8,12a,13b
\\ 
   2794	&	   ESO540-24   	&	3.98	&	*	&	  87.54	&	9.78	&	1.26	&	  *  	&	  *  	&	  *  	&	  *  	&	-0.73
&	  *  	&	6.54	&	  *  	&	  *  	&	  *  	&	  *  	&	  *  	&	  *  	&	-2.49	&	  *  	&	4.77	&	5,14b \\ 
   2796	&	   ESO540-25   	&	4.93	&	*	&	  88.92	&	10.35	&	1.42	&	  *  	&	-4.08	&	  *  	&	3.43	&	-0.78
&	  *  	&	6.73	&	  *  	&	  *  	&	  *  	&	  *  	&	  *  	&	  *  	&	-3.00	&	  *  	&	4.50	&	5,14a,14b
\\ 
   2819	&	    NGC252     	&	-1.21	&	*	&	  72.14	&	10.74	&	1.52	&	-4.14	&	  *  	&	3.55	&	  *  	&
-1.04	&	  *  	&	6.64	&	  *  	&	  *  	&	  *  	&	  *  	&	  *  	&	  *  	&	  *  	&	  *  	&	  *  	&
11,12a,13b \\ 
   2855	&	    NGC262     	&	-0.46	&	a	&	  66.07	&	10.21	&	1.38	&	-4.24	&	  *  	&	3.21	&	  *  	&	0.02
&	  *  	&	7.47	&	  *  	&	  *  	&	  *  	&	  *  	&	  *  	&	-2.37	&	  *  	&	5.08	&	  *  	&	5,12a,13b
\\ 
   2899	&	    UGC509     	&	5.89	&	*	&	  75.02	&	9.68	&	1.33	&	  *  	&	  *  	&	  *  	&	  *  	&	-0.16
&	  *  	&	6.86	&	  *  	&	  *  	&	  *  	&	  *  	&	  *  	&	  *  	&	-1.92	&	  *  	&	5.10	&	5,13b \\ 
   3019	&	    UGC524     	&	3.07	&	a	&	 155.52	&	10.82	&	1.64	&	-3.80	&	  *  	&	3.74	&	  *  	&	-0.72
&	  *  	&	6.81	&	  *  	&	  *  	&	  *  	&	  *  	&	  *  	&	-1.20	&	  *  	&	6.33	&	  *  	&	5,12a,13b
\\ 
   3051	&	    NGC278     	&	2.89	&	*	&	  11.75	&	10.01	&	0.92	&	-3.89	&	  *  	&	4.28	&	  *  	&	-0.92
&	  *  	&	7.25	&	  *  	&	-0.94	&	  *  	&	7.23	&	  *  	&	  *  	&	  *  	&	  *  	&	  *  	&
12a,13b,15,15,22 \\ 
   3055	&	    NGC279     	&	-1.25	&	*	&	  55.44	&	10.19	&	1.38	&	-3.94	&	  *  	&	3.49	&	  *  	&	  *  
&	  *  	&	  *  	&	  *  	&	  *  	&	  *  	&	  *  	&	  *  	&	  *  	&	  *  	&	  *  	&	  *  	&	11,12a \\ 
   3085	&	    NGC292     	&	8.92	&	*	&	    .08	&	9.19	&	0.87	&	-4.77	&	  *  	&	2.68	&	  *  	&	-0.24
&	  *  	&	7.21	&	  *  	&	  *  	&	  *  	&	  *  	&	  *  	&	-4.78	&	  *  	&	2.67	&	  *  	&	5,8,12a,13b
\\ 
   3343	&	    NGC312     	&	-3.82	&	*	&	 110.51	&	10.92	&	1.69	&	  *  	&	-4.70	&	  *  	&	2.83	&	  *  
&	  *  	&	  *  	&	  *  	&	  *  	&	  *  	&	  *  	&	  *  	&	  *  	&	  *  	&	  *  	&	  *  	&	11 \\ 
   3367	&	    NGC307     	&	-1.94	&	*	&	  56.47	&	10.23	&	1.42	&	  *  	&	-4.94	&	  *  	&	2.45	&	  *  
&	  *  	&	  *  	&	  *  	&	  *  	&	  *  	&	  *  	&	  *  	&	  *  	&	  *  	&	  *  	&	  *  	&	11 \\ 
   3374	&	    NGC323     	&	-4.62	&	*	&	 107.45	&	10.85	&	1.59	&	  *  	&	-4.88	&	  *  	&	2.79	&	  *  
&	  *  	&	  *  	&	  *  	&	  *  	&	  *  	&	  *  	&	  *  	&	  *  	&	  *  	&	  *  	&	  *  	&	11 \\ 
   3377	&	    NGC309     	&	4.98	&	*	&	  79.65	&	11.08	&	1.79	&	-3.76	&	  *  	&	3.74	&	  *  	&	-0.59
&	  *  	&	6.91	&	  *  	&	  *  	&	  *  	&	  *  	&	  *  	&	  *  	&	-3.60	&	  *  	&	3.90	&	5,8,12a,13b
\\ 
   3387	&	   ESO151-12   	&	-1.84	&	*	&	 102.52	&	10.51	&	1.61	&	-4.02	&	  *  	&	3.28	&	  *  	&	  *  
&	  *  	&	  *  	&	  *  	&	  *  	&	  *  	&	  *  	&	  *  	&	  *  	&	  *  	&	  *  	&	  *  	&	11 \\ 
   3434	&	    NGC311     	&	-2.00	&	*	&	  73.82	&	10.43	&	1.48	&	  *  	&	-4.56	&	  *  	&	2.91	&	  *  
&	  *  	&	  *  	&	  *  	&	  *  	&	  *  	&	  *  	&	  *  	&	  *  	&	  *  	&	  *  	&	  *  	&	11 \\ 
   3444	&	    UGC595     	&	-4.10	&	*	&	 191.25	&	10.83	&	1.73	&	  *  	&	  *  	&	  *  	&	  *  	&	  *  
&	  *  	&	  *  	&	  *  	&	  *  	&	  *  	&	  *  	&	  *  	&	  *  	&	-2.33	&	  *  	&	5.04	&	5 \\ 
   3455	&	    NGC315     	&	-4.05	&	a	&	  72.48	&	11.14	&	1.80	&	-5.87	&	  *  	&	1.65	&	  *  	&	  *  
&	-1.92	&	  *  	&	5.61	&	  *  	&	  *  	&	  *  	&	  *  	&	-3.18	&	  *  	&	4.34	&	  *  	&
2,8,11,14b,16 \\ 
   3512	&	    IC1607     	&	3.00	&	*	&	  77.13	&	10.28	&	1.31	&	-3.69	&	  *  	&	3.97	&	  *  	&	  *  
&	  *  	&	  *  	&	  *  	&	  *  	&	  *  	&	  *  	&	  *  	&	  *  	&	-2.63	&	  *  	&	5.04	&	5,14a \\ 
   3572	&	    NGC337     	&	6.72	&	*	&	  22.47	&	10.40	&	1.28	&	-4.06	&	  *  	&	3.78	&	  *  	&	-0.57
&	  *  	&	7.27	&	  *  	&	  *  	&	-1.37	&	  *  	&	6.47	&	  *  	&	  *  	&	  *  	&	  *  	&	12a,13b,22
\\ 
   3589	&	   ESO351-30   	&	-4.57	&	*	&	    .15	&	6.57	&	0.07	&	  *  	&	-5.53	&	  *  	&	0.90	&	  *  
&	  *  	&	  *  	&	  *  	&	  *  	&	  *  	&	  *  	&	  *  	&	  *  	&	-5.74	&	  *  	&	0.70	&	2,14a,16
\\ 
\multicolumn{3}{l}{continued...}					&		&		&		&		&		&		&		&		&		&		&		&		&		&		&		&		&		&		&		&		&	
\\
\hline
\end{tabular}                                              
\label{table}                                              
 } 
\end{table}                                              
\end{landscape}

\begin{table}
\caption{Number of detections (Det) and upper limit values (UL) 
available for the galaxies in Table 1. N(t) represents the total number of 
galaxies per morphological type bin.}
\tabcolsep 0.1truecm
\begin{center}
\begin{tabular}{lcrrrrrrrrrrrr}
\hline 
 	&		&  N(t) 	& \multicolumn{2}{c}{Dust}	 & & \multicolumn{2}{c}{HI} &&
\multicolumn{2}{c}{molecular}	&& \multicolumn{2}{c}{X}	\\
\cline{4-5} \cline{7-8} \cline{10-11} \cline{13-14}
Type 	&   	 t	&  	 	& 	Det	&	UL	&&	Det	&	UL	&&	Det	&   UL	&&  Det	& 	UL	\\
	&		&		&		&		&&	  	&    		&&          &   		&&    	& 		\\
\hline
E	&	-5	&	282	&	96	&	162	&&	36	&	55	&&	10	&	18	&&	87	&	73	\\
E	&	-4	&	133	&	46	&	72	&&	5	&	20	&&	2	&	1	&&	30	&	34	\\
E/S0	&	-3	&	197	&	88	&	83	&&	19	&	29	&&	8	&	3	&&	31	&	61	\\
S0	&	-2	&	307	&	154	&	114	&&	74	&	60	&&	16	&	13	&&	32	&	73	\\
S0a	&	-1	&	98	&	70	&	19	&&	41	&	22	&&	18	&	1	&&	11	&	24	\\
S0a	&	0	&	131	&	92	&	14	&&	76	&	12	&&	17	&	6	&&	12	&	27	\\
Sa	&	1	&	100	&	82	&	2	&&	67	&	9	&&	17	&	9	&&	17	&	31	\\
Sab	&	2	&	82	&	67	&	1	&&	68	&	3	&&	29	&	8	&&	14	&	36	\\
Sb	&	3	&	140	&	113	&	4	&&	117	&	3	&&	53	&	7	&&	24	&	60	\\
Sbc	&	4	&	120	&	90	&	5	&&	103	&	2	&&	48	&	4	&&	18	&	63	\\
Sc	&	5	&	126	&	98	&	5	&&	105	&	3	&&	58	&	7	&&	23	&	50	\\
Sc	&	6	&	77	&	48	&	1	&&	73	&	0	&&	20	&	11	&&	7	&	50	\\
Scd	&	7	&	40	&	25	&	0	&&	37	&	1	&&	9	&	13	&&	3	&	19	\\
Sd	&	8	&	15	&	8	&	0	&&	13	&	0	&&	3	&	2	&&	2	&	8	\\
Sm	&	9	&	29	&	13	&	0	&&	23	&	0	&&	3	&	1	&&	4	&	20	\\
Irr	&	10	&	39	&	14	&	1	&&	34	&	1	&&	6	&	6	&&	4	&	25	\\
  &    &    &    &    &&    &    &&    &    &&    &    \\
\hline                                              
Total	&		&	1916	&	1104	&	483	&&	891	&	220	&&	317	&	110	&&	319	&	654	\\
  &    &    &    &    &&    &    &&    &    &&    &    \\
\end{tabular}
\label{numbers}
\end{center}
\end{table}

\begin{table}                                              
\caption{Mean warm dust content according to morphological type. N represents the 
total number of galaxies which have data (detections and upper limits), while N$_{ds}$ 
indicates the number of detections resulting from the survival analysis 
statistics. When upper limits exist, mean values are 
derived making use of survival analysis. All values are normalized to \LB, the 
corrected blue luminosities, and to \area, the square of the diameter in kpc 
at the isophote of 25 \sbu.}                                              
\begin{center}
\begin{tabular}{lrrrr}                                              
\hline                                              
Type  &  t  &  N/N$_{ds}$ &  \md/\LB         &  \md/\area      \\
      &     &          &          &         \\
\hline                                              
E	&	-5	&	259/98 &	-5.99	$\pm$	0.12	&	1.53	$\pm$	0.13	\\	
E	&	-4	&	118/47 &	-5.60	$\pm$	0.13	&	1.94	$\pm$	0.14	\\	
E/S0	&	-3	&	171/89 &	-5.19	$\pm$	0.09	&	2.42	$\pm$	0.09	\\	
S0	&	-2	&	269/156 &	-4.97	$\pm$	0.07	&	2.51	$\pm$	0.08	\\
S0a	&	-1	&	87/69	&	-4.63	$\pm$	0.13	&	2.79	$\pm$	0.12	\\
S0a	&	0	&	108/95 &	-4.34	$\pm$	0.08	&	3.10	$\pm$	0.08	\\
Sa	&	1	&	82/80	&	-4.07	$\pm$	0.06	&	3.49	$\pm$	0.06	\\
Sab	&	2	&	68/67	&	-3.82	$\pm$	0.05	&	3.70	$\pm$	0.05	\\
Sb	&	3	&	117/113 &	-3.75	$\pm$	0.04	&	3.83	$\pm$	0.05	\\
Sbc	&	4	&	96/91	&	-3.74	$\pm$	0.03	&	3.83	$\pm$	0.04	\\
Sc	&	5	&	102/97 &	-3.78	$\pm$	0.03	&	3.82	$\pm$	0.04	\\
Sc	&	6	&	50/49	&	-3.92	$\pm$	0.05	&	3.58	$\pm$	0.07	\\
Scd	&	7	&	24/24	&	-4.09	$\pm$	0.07	&	3.34	$\pm$	0.09	\\
Sd	&	8	&	8/8	&	-4.50	$\pm$	0.23	&	3.13	$\pm$	0.24	\\
Sm	&	9	&	13/13	&	-4.40	$\pm$	0.21	&	3.04	$\pm$	0.24	\\
Irr	&	10	&	15/14	&	-4.12	$\pm$	0.16	&	3.43	$\pm$	0.19	\\
  &    &        &        &        \\
\hline                                              
Total	&		&	1587/1110 &  &	\\					
\end{tabular}                                              
\end{center}
\label{Tdust}                                              
\end{table}

\begin{table}                                              
\caption{Mean HI content according to morphological type. Notation as in Table \ref{Tdust}.}                                              
\begin{center}
\begin{tabular}{lrrrr}                                              
\hline                                              
Type  &  t  &  N/N$_{ds}$  &  \mhi/\LB      &  \mhi/\area    \\
      &     &           &        &        \\
\hline                                              
E	&	-5	&	91/37	 &	-2.34	$\pm$	0.14	&	5.33	$\pm$	0.13	\\
E	&	-4	&	25/6 	 &	-1.83	$\pm$	0.12	&	5.54	$\pm$	0.14	\\
E/S0	&	-3	&	49/20	 &	-1.89	$\pm$	0.16	&	5.70	$\pm$	0.15	\\
S0	&	-2	&	134/75 &	-1.55	$\pm$	0.09	&	5.96	$\pm$	0.09	\\
S0a	&	-1	&	62/41	&	-1.81	$\pm$	0.19	&	5.58	$\pm$	0.19	\\
S0a	&	0	&	88/77	&	-0.96	$\pm$	0.06	&	6.47	$\pm$	0.06	\\
Sa	&	1	&	76/67	&	-1.05	$\pm$	0.07	&	6.52	$\pm$	0.07	\\
Sab	&	2	&	71/68	&	-0.95	$\pm$	0.06	&	6.57	$\pm$	0.06	\\
Sb	&	3	&	120/117 &	-0.77	$\pm$	0.04	&	6.78	$\pm$	0.03	\\
Sbc	&	4	&	106/104 &	-0.66	$\pm$	0.03	&	6.90	$\pm$	0.03	\\
Sc	&	5	&	107/104 &	-0.67	$\pm$	0.04	&	6.92	$\pm$	0.03	\\
Sc	&	6	&	74/74	&	-0.48	$\pm$	0.04	&	6.93	$\pm$	0.04	\\
Scd	&	7	&	37/36	&	-0.33	$\pm$	0.07	&	6.96	$\pm$	0.05	\\
Sd	&	8	&	13/13	&	-0.32	$\pm$	0.19	&	7.11	$\pm$	0.11	\\
Sm	&	9	&	23/23	&	-0.29	$\pm$	0.08	&	7.02	$\pm$	0.06	\\
Irr	&	10	&	35/34	&	-0.20	$\pm$	0.10	&	7.05	$\pm$	0.06	\\
      &     &           &        &        \\
\hline                                              
Total	&		&	1111/	896	&		&		\\
\end{tabular}                                              
\end{center}
\label{Thi}                                              
\end{table}

\begin{table}																							
\caption{Mean content of molecular gas according to morphological type. Notation as in Table
\ref{Tdust}. The values for t=-4 and t=8 are based on a low number of galaxies and are more
uncertain.}																							
\begin{center}
\begin{tabular}{lrrrr}																							
\hline																							
Type  &  t     & N/N$_{ds}$ &  \mmol/\LB       &  \mmol/\area    \\
      &        &         &          &         \\
\hline                                              
E	&	-5	&	28/11	&	-2.53	$\pm$	0.22	&	5.03	$\pm$	0.22	\\
E	&	-4	&	3/3	&	-1.25	$\pm$	0.23	&	6.34	$\pm$	0.22	\\
E/S0	&	-3	&	11/8	&	-2.23	$\pm$	0.32	&	5.33	$\pm$	0.33	\\
S0	&	-2	&	30/17	&	-1.82	$\pm$	0.21	&	5.76	$\pm$	0.22	\\
S0a	&	-1	&	18/18	&	-1.41	$\pm$	0.19	&	6.00	$\pm$	0.17	\\
S0a	&	0	&	24/18	&	-1.14	$\pm$	0.15	&	6.33	$\pm$	0.18	\\
Sa	&	1	&	25/18	&	-1.13	$\pm$	0.17	&	6.52	$\pm$	0.16	\\
Sab	&	2	&	37/30	&	-1.17	$\pm$	0.13	&	6.39	$\pm$	0.16	\\
Sb	&	3	&	60/53	&	-0.96	$\pm$	0.08	&	6.63	$\pm$	0.08	\\
Sbc	&	4	&	53/49	&	-0.88	$\pm$	0.06	&	6.75	$\pm$	0.07	\\
Sc	&	5	&	64/58	&	-1.01	$\pm$	0.06	&	6.58	$\pm$	0.07	\\
Sc	&	6	&	31/20	&	-1.44	$\pm$	0.11	&	5.99	$\pm$	0.14	\\
Scd	&	7	&	22/9	&	-1.92	$\pm$	0.16	&	5.54	$\pm$	0.18	\\
Sd	&	8	&	5/3	&	-1.52	$\pm$	0.35	&	6.28	$\pm$	0.34	\\
Sm	&	9	&	4/4	&	-1.54	$\pm$	0.32	&	6.05	$\pm$	0.49	\\
Irr	&	10	&	12/6	&	-1.71	$\pm$	0.28	&	5.90	$\pm$	0.35	\\
      &        &         &          &         \\
\hline                                              
Total	&		&	427/325	&		&		\\
\end{tabular}                                              
\end{center}
\label{Tmol}                                              
\end{table}

\begin{table}                                              
\caption{Mean X-ray luminosity of the galaxy \Lx\ versus the morphological type t.
All values are normalized to \LB, the corrected blue luminosities, or \area, the 
square of the diameter in kpc at the isophote of 25 \sbu. The values of the first table 
were calculated excluding the galaxies hosting AGN. The second table includes them.}                                              
\begin{center}
\begin{tabular}{lrrrr}                                              
\hline
Without AGN \\
\hline
Type	&	t	&	N/N$_{ds}$	&	Log L$_{Xt}$/L$_B$	&	Log L$_{Xt}$/D$^2_{25}$	\\
\hline
E	&	-5	&	150/79	&	-3.69	$\pm$	0.122	&	3.62	$\pm$	0.19	\\
E	&	-4	&	61/29	&	-3.42	$\pm$	0.14	&	4.08	$\pm$	0.15	\\
E/S0	&	-3	&	92/31	&	-3.97	$\pm$	0.22	&	3.72	$\pm$	0.16	\\
S0	&	-2	&	91/24	&	-3.78	$\pm$	0.10	&	3.61	$\pm$	0.11	\\
S0a	&	-1	&	28/7	&	-3.82	$\pm$	0.07	&	3.68	$\pm$	0.09	\\
S0a	&	0	&	31/4	&	-3.67	$\pm$	0.08	&	3.69	$\pm$	0.11	\\
Sa	&	1	&	38/8	&	-4.03	$\pm$	0.32	&	3.75	$\pm$	0.15	\\
Sab	&	2	&	40/7	&	-3.99	$\pm$	0.14	&	3.47	$\pm$	0.12	\\
Sb	&	3	&	63/10	&	-4.24	$\pm$	0.15	&	3.32	$\pm$	0.13	\\
Sbc	&	4	&	65/8	&	-4.01	$\pm$	0.08	&	3.49	$\pm$	0.09	\\
Sc	&	5	&	65/19	&	-4.04	$\pm$	0.09	&	3.50	$\pm$	0.11	\\
Sc	&	6	&	57/7	&	-4.29	$\pm$	0.12	&	3.01	$\pm$	0.17	\\
Scd	&	7	&	20/2	&	-4.26	$\pm$	0.03	&	3.16	$\pm$	0.08	\\
Sd	&	8	&	9/1	&	-5.17	$\pm$	0.03	&	1.83	$\pm$	0.00	\\
Sm	&	9	&	23/4	&	-4.22	$\pm$	0.28	&	3.06	$\pm$	0.29	\\

Irr	&	10	&	28/3	&	-3.85	$\pm$	0.23	&	3.43	$\pm$	0.17	\\
      &        &         &          &         \\
\hline                                              
Total	&		&	861/243	&			&		\\
\hline
With AGN \\
\hline
Type	&	t	&	N/N$_{ds}$	&	Log L$_{Xt}$/L$_B$	&	Log L$_{Xt}$/D$^2_{25}$	\\
\hline
Total	&		&	945/280	&				&				\\
E & -5	&	161/88 &	-3.65	$\pm$	0.12	&	3.65	$\pm$	0.18	\\
E & -4	&	63/31	&	-3.36	$\pm$	0.14	&	4.15	$\pm$	0.15	\\
E/S0 & -3	&	94/32	&	-4.03	$\pm$	0.22	&	3.69	$\pm$	0.16	\\
S0 & -2	&	104/33 &	-3.59	$\pm$	0.11	&	3.82	$\pm$	0.12	\\
S0/a & -1	&	35/11	&	-3.60	$\pm$	0.15	&	3.89	$\pm$	0.16	\\
S0/a & 0	&	39/12	&	-3.30	$\pm$	0.14	&	4.11	$\pm$	0.16	\\
Sa & 1	&	48/17	&	-3.69	$\pm$	0.25	&	4.09	$\pm$	0.17	\\
Sab & 2	&	50/15	&	-3.79	$\pm$	0.13	&	3.72	$\pm$	0.12	\\
Sb & 3	&	84/25	&	-3.98	$\pm$	0.12	&	3.57	$\pm$	0.12	\\
Sbc & 4	&	81/18	&	-3.82	$\pm$	0.09	&	3.69	$\pm$	0.09	\\
Sc & 5	&	73/24	&	-3.97	$\pm$	0.09	&	3.57	$\pm$	0.10	\\
Sc & 6	&	58/7	&	-4.29	$\pm$	0.12	&	3.01	$\pm$	0.17	\\
Scd & 7	&	21/3	&	-4.20	$\pm$	0.06	&	3.23	$\pm$	0.09	\\
Sd & 8	&	10/2	&	-4.91	$\pm$	0.18	&	2.60	$\pm$	0.54	\\
Sm & 9	&	24/4	&	-4.28	$\pm$	0.24	&	3.01	$\pm$	0.26	\\
Irr & 10	&	29/4	&	-3.88	$\pm$	0.17	&	3.40	$\pm$	0.15	\\
      &        &         &          &         \\
\hline                                              
Total & 	&	974/326	&		&			\\
\end{tabular}                                              
\end{center}
\label{LX_t}                                              
\end{table}                                              

\clearpage

\clearpage

\end{document}